# Observation of negative refraction of Dirac fermions in graphene

Gil-Ho Lee[†], Geon-Hyoung Park and Hu-Jong Lee[*]

Department of Physics, Pohang University of Science and Technology, Pohang 790-784, Republic of Korea

[†]Current address: Department of Physics, Harvard University, Cambridge, Massachusetts 02138, USA

[*]Correspondence and requests for materials should be addressed to H.-J.L. (email: hjlee@postech.ac.kr).

**Half a century ago, Veselago[1] proposed 'left-handed' materials with negative permittivity and permeability, in which waves propagate with phase and group velocities in opposite directions. Significant work has been undertaken to attain this left-handed response, such as establishing a negative refractive index in so-called metamaterials, which consist of periodic sub-wavelength structures[2-4]. However, an electronic counterpart has not been demonstrated owing to difficulties in creating repeated structures smaller than the electronic Fermi wavelength ($\lambda_F$) of the order ~ 10 nm. Here, without needing to engineer sub-wavelength structures, we demonstrate negative refractive behaviour of Dirac fermions in graphene, exploiting its unique relativistic band structure[5]. Analysis of both electron focusing through a n-p-n flat lens and negative refraction across n-p junctions confirms left-handed behaviour in the electronic system. This new approach to electronic optics is of particular relevance to the on-going efforts to develop novel quantum devices with emerging[6] layered materials.**

Due to their wave nature, electrons follow the laws of optics when their mean free path and phase coherence length are larger than the system size. To date, electron optics has been demonstrated mostly in conventional two-dimensional electron gas (2DEG) systems such as interferometers or electrostatic lenses[7]. Graphene can provide an attractive platform for studying the unique electronic optics of Dirac fermions owing to its gapless and linear dispersion. Cheianov *et al.*[5] proposed the interesting idea that transparent ballistic n-p junctions of graphene can exhibit negative refractive behaviour with electrostatic gates providing control of local doping (Fig. 1a). This is a fundamentally different approach from the conventional one utilising periodic sub-wavelength structures in metamaterials. In this approach, negative refraction is observed because the wave vector ($k$) and the group velocity [$v_g = dE(k)/d(\hbar k)$] of carriers are parallel or antiparallel to each other depending on whether the carriers are electron-like [$E(k) = \hbar v k$] or hole-like [$E(k) = -\hbar v k$], respectively. When an electronic wave enters an n-p junction, the tangential component of $v_g$ is reversed to conserve the tangential component of $k$ ($k_i \sin\theta_i = -k_r \sin\theta_r$), while the perpendicular component of $v_g$ itself is preserved (Fig. 1b). As a consequence, the refraction follows Snell's law with negative reflective index $n$, where

$$n \equiv \frac{\sin\theta_i}{\sin\theta_r} = -\frac{k_r}{k_i}.$$  (Eq. 1)

Here, $n$ is defined as the relative refractive index of the p-doped region compared to the n-doped region. A vanishing bandgap, a distinct property of graphene, is an essential component of the effect, since it facilitates transparent p-n junctions for electron optics; in contrast a semiconducting 2DEG results in impermeable p-n junctions due to the depletion region. Although graphene p-n junctions have been extensively studied[8-15] with clear evidence of Klein tunnelling[16,17], negative refraction has not been confirmed clearly due to either the diffusive nature of the carriers or because the p-n potential barrier was smoother than $\lambda_F$ ( $= 2\pi/k$ ~ a few tens of nm).

We fabricated a graphene-based flat lens device[1,18] (or a Veselago lens) consisting of two successive n-p and p-n junctions as shown in Figs. 1c and d. When the focusing condition ($n = -b/a$) is met, electron beams from the IN port are negatively refracted successively through two n-p and p-n junctions and then refocused at the OUT port. Monolayer graphene was encapsulated in insulating and atomically flat boron nitride (BN) crystals to achieve the characteristic ballistic nature of graphene (see Method). The top local gate voltage ($V_T'$) and bottom global gate voltage ($V_B$) enabled in situ control of the refractive index. A thin top BN layer (thickness ~ 14 nm) provided sharp p-n junctions with a characteristic sharpness of $d$ ~ 12 nm (see Supplementary Information).

The geometric asymmetry ($a = 600$ nm, $b = 720$ nm) is exaggerated in the schematic shown in Fig. 1c. Each corner with a narrow constriction (width $w$ ~ 100 nm) was connected to the Cr/Au metal electrode. In situ etching of the graphene just before metal deposition was implemented to make a highly transparent contact. This geometry led to a negative van der Pauw resistivity at temperature $T = 4.2$ K, confirming ballistic transport (see Supplementary Information). A bias current was injected ($I_{bias} = 500$ nA), and the drain currents at the other three ports were measured simultaneously as a function of $V_T'$ and $V_B$. The compensated top gate voltage $V_T = \gamma V_T' + V_B$, and $V_B$ determined the carrier densities $\rho_T = k^2/\pi = \eta V_T$ and $\rho_B = \eta V_B$ of the top- and back-gated regions, respectively, with $\eta$ ~ $7.2\times10^{-2}$ cm$^{-2}$V$^{-1}$. The ratio of top to bottom gate efficiency, $\gamma = 21.0$, was determined using $V_T'$ and $V_B$ of Dirac points. Independent control of $V_T$ and $V_B$ allowed exploration of a range of values of $n = \text{sgn}(V_T \cdot V_B)\sqrt{V_T/V_B}$ for both polarities. The drain current at port 2, $I_2$, taken at 100 K is plotted as a function of $V_T$ in Fig. 2a. Near $V_T = 0$, where the top-gated region becomes most resistive, a greater current tends to flow toward ports 1 or 3 rather than port 2, resulting in a decrease of $I_2$ at all values of $V_B$ (see Supplementary Information). On top of the background, each trace for $V_T$ for different values of $V_B$ exhibits a current focusing peak (red triangles) only in the bipolar regimes of n-p-n and p-n-p. As $V_B$ moves away from the Dirac point, the focusing peaks become smaller and broader, eventually obscured by the background signal. This happens as $\lambda_F$ becomes much shorter than $d$ for $V_{B,T}$ away from the Dirac point, resulting in more electron reflection at the p-n boundaries[19,20]. After subtracting the data in this regime as a background signal, the gate dependence of the peak positions and widths becomes more conspicuous (Fig. 2b). Current focusing peaks shift linearly with $V_B$ for negative values of $V_T/V_B$. This confirms that they originate from electron focusing through the p-n boundaries with a negative value of $n$, rather than any sample-specific doping inhomogeneity. The focusing condition is estimated to be $n = -b/a = -1.20$ for our device geometry. This value agrees well with the observations: $n_{npn} = -1.35$ (for n-p-n) and $n_{pnp} = -1.20$ (for p-n-p), as shown in Fig. 2c. The solid red line represents the numerical simulation result of a classical particle tracing with geometric parameters; $a$, $b$, $w$, and $d$ (see Supplementary Information) with the angle dependence of the transmission probability $T$ through p-n boundaries[20],

$$T(n,\theta_i) = \frac{\cos\theta_i \cos\theta_r}{\cos^2\left[(\theta_i+\theta_r)/2\right]} \exp\left(-\pi k_F d \sin^2\theta_i \frac{2}{1-n}\right), \quad \text{(Eq. 2)}$$

and the Fermi velocity of graphene $v_F \sim 10^6$ ms$^{-1}$. For the unipolar case, the exponential term in Eq. 2 can be ignored because there is no Wentzel–Kramers–Brillouin (WKB) tunnelling term. The $V_T/V_B$ slope of the simulation result also agrees well with the data. A finite offset (~ –2.8 V) in $V_T$ for the linear fitting in the n-p-n regime is as small as the full-width-half-maximum (FWHM) of the Dirac point ($\Delta V \sim 2.5$ V), which represents the inhomogeneity of the Fermi level. Thus, the offset may have resulted from ambiguity in the Dirac point.

The width of the current focusing peaks is related to the edge roughness of the top gate. The specularity of a p-n boundary was quantified by adopting the empirical Phong's model[21]. In this model, $\alpha$ parameterizes the specularity of the interface with a distribution probability proportional to $\cos^\alpha \theta$, where the refracted beam deviates from the refraction angle of the perfect specularity case by an error angle $\theta$. In Fig. 2d, the simulation with $\alpha =$ 20 (FWHM of the probability distribution ~ 30°) shows significant resemblance to the observed results of the p-n-p case for both the width and asymmetricity of the $I$ vs. $V_T$ curves. The simulation for infinite $\alpha$ (perfect specularity, FWHM of distribution = 0°) shows much larger and sharper focusing peaks, implying that flatter p-n boundaries can significantly improve Veselago focusing.

The temperature dependence of the Veselago focusing and the quantum interference survived up to 100 K and 40 K, respectively (Fig. 3a). Temperature dependences of both features are more pronounced in a plot of the numerical derivative $dI_2/dV_T$ (Fig. 3b). Since focusing relies on classical electron trajectories, it can persist at higher temperatures than quantum interference. A simple length scale analysis shows that the focusing signal can persist up to a temperature of an order $\hbar v_F k_F(w/c) \sim 170$ K, where the spread of the perpendicular component of momentum ($\Delta k \sim T/\hbar v_F$) becomes sufficiently wide that electrons fail to reach the constricted detector. Here, $c = (a+b)/2$ is the characteristic propagation length. In addition, the ballistic nature is weakened by electron-phonon scatterings. We therefore believe that the observed temperature dependence of the focusing peaks results from the thermally broadened momentum distribution and the reduced mean free path.

Below 40 K, quasi-periodic oscillations appear, increasing near the focusing peak with a period of ~ 2 V [Fig. 3a]. These oscillations are likely a result of quantum interference of refracted electron beams with different travel distances[5]. This results in spatial modulation of the current density near the focal point with a period of $2\lambda_F$ for $n$ ~ -1.2. Given that the focal point moves by an amount $\Delta\delta = (\partial\delta/\partial V_T)\Delta V_T = -b\sqrt{V_B/V_T^3}\Delta V_T$ with changing $\Delta V_T$, the oscillation period $\Delta V_T = 2$ ~ 3 V is estimated under the condition $\Delta\delta \sim 2\lambda_F$. Here, $\delta = 2(b/|n|-a)$ is the focal position with respect to the detection constriction. According to the simple energy scale analysis, phase coherent oscillation survives up to a temperature of an order $\hbar v_F/c$ ~ 10 K, which is reasonably consistent with the observed temperature dependence. The data in Fig. 2a was taken at a finite temperature of $T = 100$ K to study electron focusing behaviour while suppressing this quantum interference. Another possible origin of the oscillation is a Fabry-Perot-type interference phenomenon of successively reflected electrons in the cavity of the top-gated region. However, the expected Fabry-Perot period $\Delta V_T$ ~ 0.4 V with a cavity size of $2b$ deviates significantly from the measured one. Compared to a previous study with a diffusive back-gated region and wide cavity geometry[17], the ballistic nature of the back-gated region and the narrow cavity of the Veselago lens in this study supplied more electrons along the normal incidence; the consequent lack of backscattering of those electrons leads to a less pronounced Fabry-Perot resonance. By applying a magnetic field, however, the Fabry-Perot oscillation can be enhanced, the period of which matches well with the expected value (see Supplementary Information).

Thus far, current measurements for verification of the electronic Veselago focusing through the graphene heterostructures have been discussed. To improve differentiation of the Veselago negative refraction signal from the large background, a voltage measurement scheme in a non-local geometry was adopted. Previously, this technique has been adopted to study graphene phenomena such as spin transport[22], the spin Hall effect[23] or more recently the valley Hall effect[24]. Our non-local device consisted of a single p-n junction in circular ballistic graphene with multiple directional leads (Fig. 4a). Non-local resistance, $R_{nl} = (V^+ - V^-)/I_{bias}$, was measured as a function of $V_T$ and $V_B$. When two semicircular graphene layers had similar carrier density with the same polarity ($V_T \sim V_B$), injected electrons (black arrow) were refracted with positive $n$ (~ 1) and reached the electrode $V^+$ (black dotted arrow). Electrons accumulated at the electrode $V^+$ lowered the potential and resulted in a negative $R_{nl}$ ($V^+ < V^-$; blue-coloured regions in Fig. 4b), confirming the ballistic nature of the system at $T \leq \sim 60$ K. In contrast, in a bipolar regime with a similar carrier density ($V_T \sim -V_B$) the p-n junction deflected injected electrons toward the electrode $V^-$ with a negative $n$ (~ –1; red dotted arrow), which reversed the sign of $R_{nl}$ to positive. Directive long guiding constrictions set the incident angle ($\theta_i = 21°$) for the injected electrons and the refraction angle ($\theta_r = -23°$) for the electrons to be refracted toward the electrode $V^-$. This gives a focusing condition of $V_T/V_B = -0.84$ or equivalently $n = -0.92$ represented by the dotted white line in Fig. 4b, along which $R_{nl}$ is most strongly enhanced. In Fig. 4c, line cuts of $R_{nl}$ as a function of $V_T$ at a fixed $V_B$ clearly show the enhancement of $R_{nl}$ for $T \leq 60$ K, although small fluctuations are present through superposition of the quantum interference signal. A similar length scale analysis to that used above suggests that the focusing signal should persist up to ~ 200 K. However, the coincident disappearance of the enhanced $R_{nl}$ in the bipolar regime and the negative $R_{nl}$ in the unipolar regime above 100 K suggests that the focusing effect was limited by the ballisticity of the system.

Finally, we comment on the reproducibility of the results discussed above as well as on perspectives for improving this phenomenon. Two additional devices (one with current measurement and the other with non-local voltage measurement) reproduced essentially the same negative refraction behaviour discussed so far (see Supplementary Information).

Ballistic graphene heterostructures with sharp p-n junctions enabled investigation of unique electronic optics with gate-tuneable negative refractive index. One can improve the quality of graphene electronic optics by adopting readily available fabrication technologies, such as a superlattice top gate to collimate the injected electron beam[25], a quantum point contact for the detector constriction[26], or mechanically cleaved graphite as a top gate electrode with atomically flat edges. Absorbers[7] around the device edges also can help eliminate the parasitic electrons and enhance focusing visibility. These efforts will lead to other interesting physics such as Klein tunneling[17], specular Andreev reflection with superconducting contacts[27,28], or a Cooper pair splitter[29,30]. With a long mean free path (~ 1 μm) at room temperature, due to very weak electron-phonon interactions at a large optical phonon energy, graphene promises novel components for electronic optics operating at high temperatures.

**Methods**

**Sample fabrication.** BN was mechanically exfoliated on polypropylene carbonate (PPC) spun on a silicon (Si) substrate, peeled off and transferred to a Gel film (Gel-Pak, PF-30/17-X4) prepared on a glass slide. A BN/PPC/Gel-film stamp was used to make a stacked BN/graphene (G)/BN structure on an Si substrate covered with 300 nm $SiO_2$ by successively aligning and picking up the graphene and the basal BN[31]. BN/G/BN stacks were etched using $CF_4$ reactive ion etching using PMMA (Poly-methyl methacrylate) polymer as an etching mask. To improve graphene edge contacts with the metal electrode, we used the same PMMA polymer layer as an etching mask as well as a lift-off resist layer such that a freshly etched graphene edge is never contaminated by the polymer. To connect the top gate to the outer electrode, exposed graphene edges were passivated with a 50-nm-thick aluminium oxide layer, which prevents electrical shortage between the top gate electrode and the exposed graphene edge (see Supplementary Information).

**Supplementary Information** available online.

**Acknowledgments**

We thank M. Kim, J. H. Lee, and J. Lee for helpful discussions on the device fabrication. We thank R. N. Sajjad and A. W. Ghosh for the fruitful discussions and critical reading the manuscript. We also thank P. Kim for the valuable comments. This work was supported by the National Research Foundation (NRF) through the SRC Center for Topological Matter (Grant No. 2011-0030046) and the GFR Center for Advanced Soft Electronics (Grant No. 2014M3A6A5060956).

**Author contribution**

G.-H.L. and H.-J.L. conceived the idea and designed the project. H.-J.L. supervised the project. G.-H.L. and G.-H.P. fabricated the devices. G.-H.L. performed the measurements. G.-H.L. and H.-J.L. analysed the data and wrote the manuscript.

**Figures and captions**

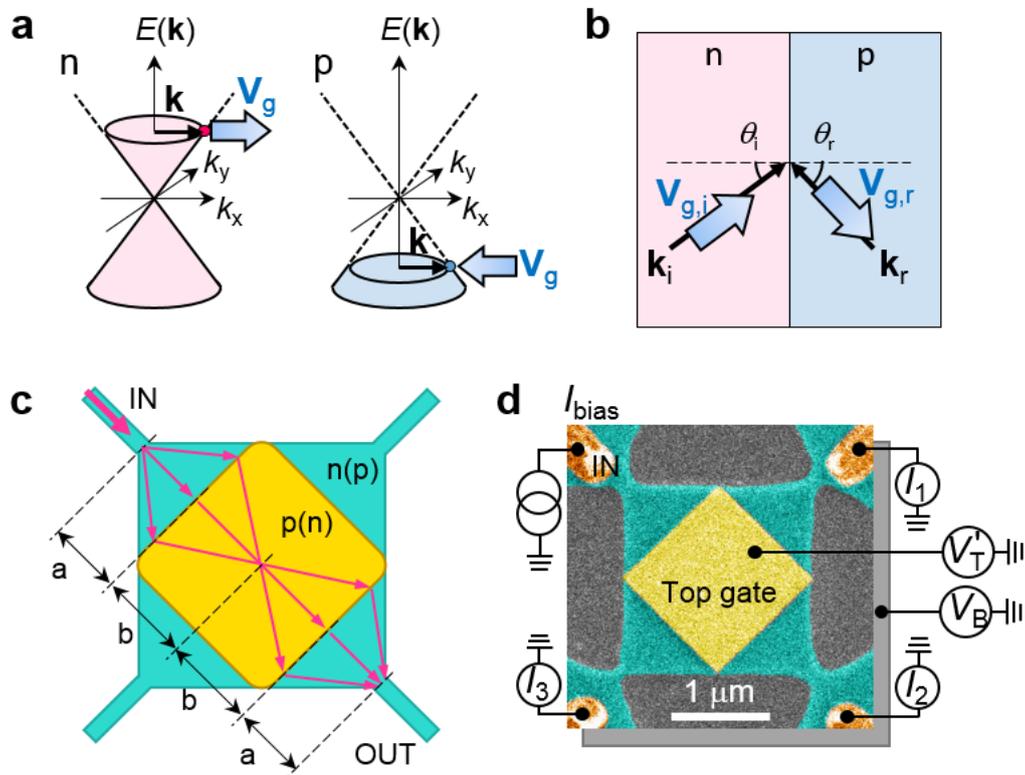

**Figure 1 | Negative refraction in graphene p-n junctions. a**, Band structure of graphene and the dispersion relation of the wave vector (**k**) and the group velocity (**V**$_g$) for the electron-doped (n) and hole-doped (p) states. Red (blue) circles represent electron-(hole-)like quasiparticles. **b**, Negative refraction across the n-p junction. **c**, Schematic of Veselago lens with current trajectories under a focusing condition, where spreading electrons from the port IN are refocused on port 2. **d**, Scanning electron microscopy image of the Veselago lens device taken before attaching top gate bridge connection. Orange, turquoise, and yellow colours represent the Cr/Au electrodes, the boron nitride (BN)/graphene/BN stack, and the Cr/Au top gate, respectively.

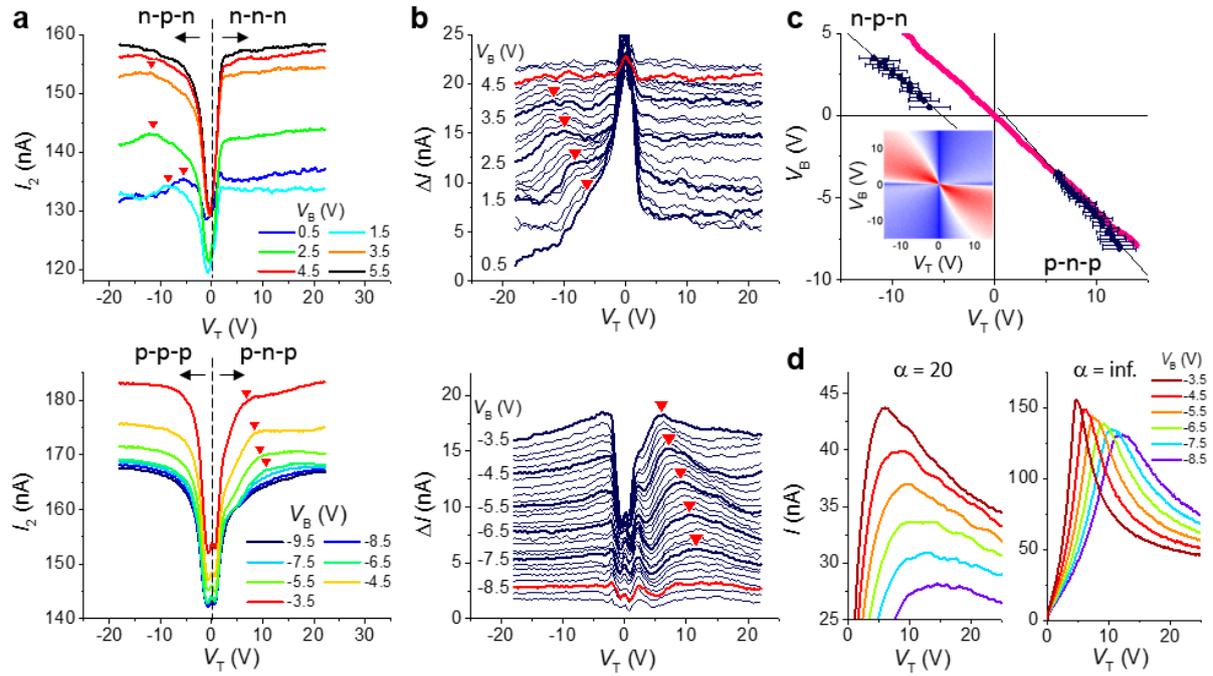

**Figure 2 | Current focusing of graphene-based Veselago lens. a**, $V_T$ dependence of $I_2$ with positive (upper panel) and negative (lower panel) values of $V_B$ at temperature $T = 100$ K. Red inverted triangles indicate the current enhancement peak of each trace. Black thicker lines represent the background without the current enhancement peak. **b**, Background-subtracted current $\Delta I = I_2 - I_2$ ($V_B = 5.5$ for $V_B > 0$; upper panel, or $V_B = -9.5$ V for $V_B < 0$; lower panel) as a function of $V_T$. Traces are separated in steps of $V_B = 0.2$ V with an arbitrary offset for clarity. Every fifth trace is emphasised with a thicker line. Red triangles indicate the current enhancement peaks and red lines are the boundaries beyond which current enhancement vanishes. **c**, Relationship between $V_T$ and $V_B$ of the current enhancement peaks (symbols) with linear fitting lines (black lines). Error bars denote the uncertainty of the peak position due to the fluctuation of $\Delta I$. The solid red line represents the simulation result. Inset, simulated focusing current with a colour scale from blue (0 nA) to red (40 nA). **d**, Simulation of the focusing current, $I$, for $V_B < 0$ with specularity parameter $\alpha = 20$ (left panel) and infinite $\alpha$ (right panel). In the left panel, each curve was offset by a step of 2 nA from bottom to top for clear comparison with the lower panel of b.

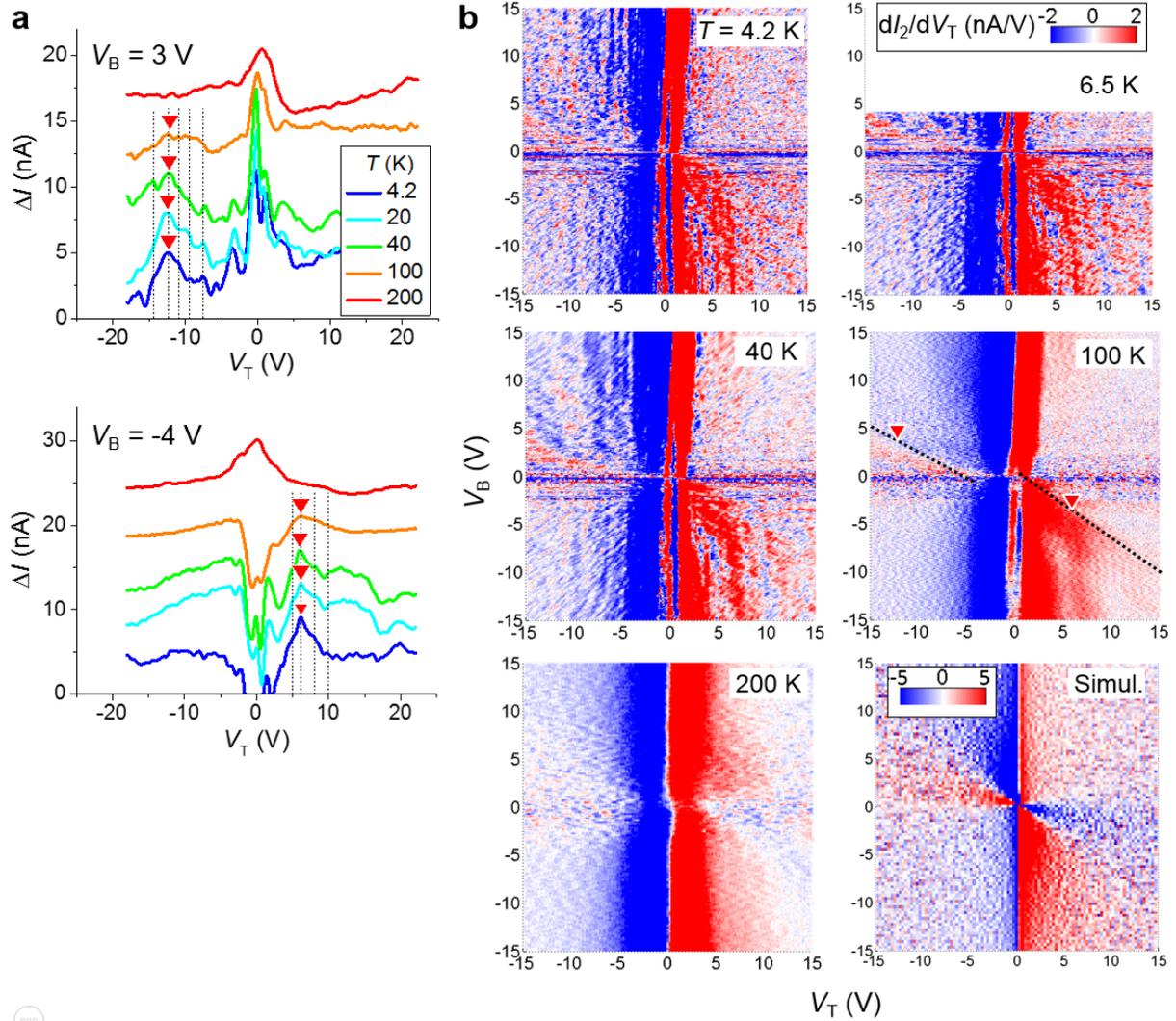

**Figure 3 | Temperature dependence of current focusing. a**, Background-subtracted current $\Delta I$ as a function of $V_T$ for $V_B = 3$ V (upper panel) and $V_B = -4$ V (lower panel) taken at various temperatures ($T$). Red triangles indicate the current enhancement peaks, which gradually vanish as $T$ increases. **b**, Colour plot of $dI_2/dV_T$ as a function of $V_T$ and $V_B$ taken at $T = 4.2$ K, 6.5 K, 40 K, 100 K, and 200 K. The last panel is the simulation result. Linear fitting lines obtained in Fig. 2d are overlaid on the $T = 100$ K plot as dotted lines. They represent the boundaries between positive (red) and negative values (blue) of $dI_2/dV_T$, which correspond to the peaks of $I_2$.

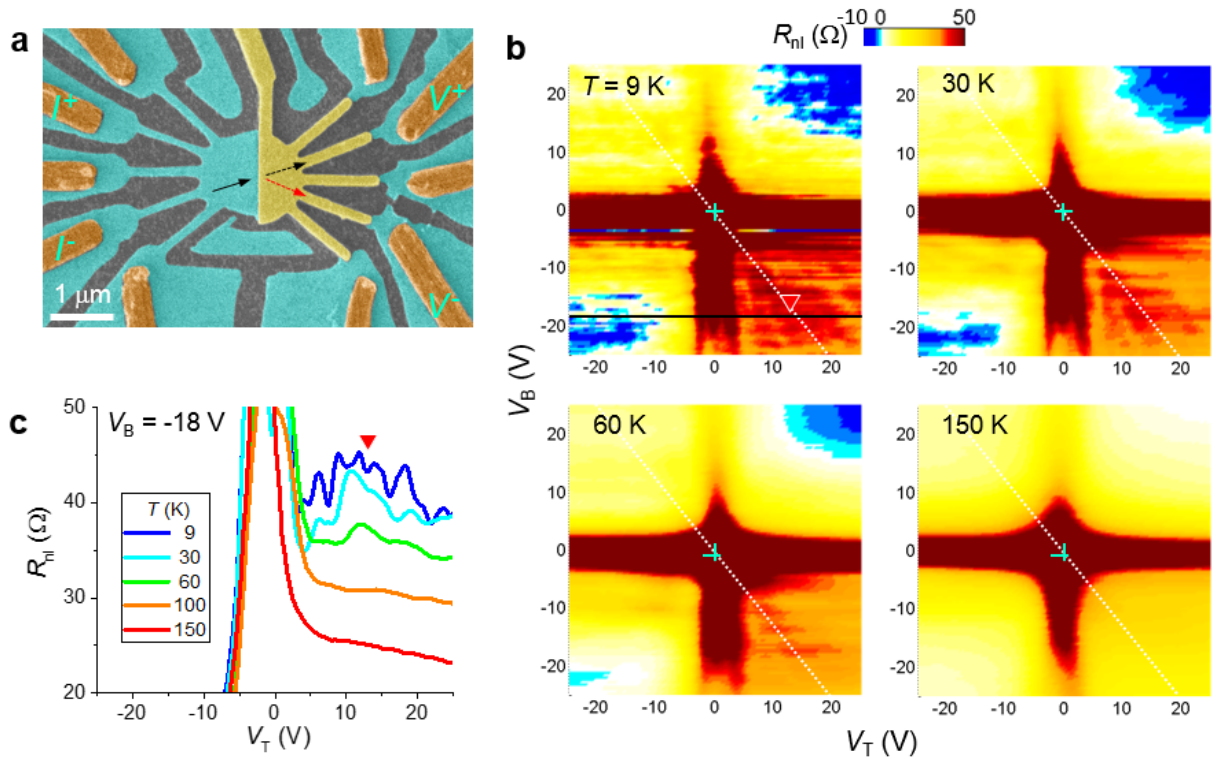

**Figure 4 | Voltage measurement for negative refraction. a**, Scanning electron microscopy (SEM) image of a non-local measurement device. While injecting a bias current ($I_{bias}$) from $I^+$ to $I^-$, a voltage difference $\Delta V = V^+ - V^-$ was measured. Injected electrons (black solid arrow) can be refracted with either positive (black dotted arrow) or negative (red dotted arrow) refractive index depending on the top ($V_T$) and bottom gate voltages ($V_B$). **b**, Quadrant map of non-local resistance, $R_{nl} = \Delta V / I_{bias}$, at $T = 9$ K, 30 K, 60 K, and 150 K. The expected focusing condition is represented by a white dotted line. **c**, $V_T$ dependence of $R_{nl}$ at $V_B = -18.4$ V (black line in b) at various temperatures. The inverse triangle represents the position of $R_{nl}$ peaks.